\documentclass[aps,pre,onecolumn,floatfix]{revtex4}

\usepackage{graphicx}
\usepackage{amssymb,amsfonts,amsmath}
\usepackage{mathtools}
\usepackage{epsf}
\usepackage{subfigure}
\usepackage{epstopdf}

\newcommand{\be}{\begin{equation}}
\newcommand{\ee}{\end{equation}}
\newcommand{\bea}{\begin{eqnarray}}
\newcommand{\eea}{\end{eqnarray}}

\begin{document}

\title{\bf Chargaff's second parity rule and the kinetics of DNA replication}

\author{Pierre Gaspard}
\affiliation{Center for Nonlinear Phenomena and Complex Systems,\\
Universit\'e Libre de Bruxelles (ULB), Code Postal 231, Campus Plaine,
B-1050 Brussels, Belgium}

\begin{abstract}
This paper presents the study of a DNA replication model grounded in the biochemical kinetics of DNA polymerases, which copy each DNA strand into a complementary strand, except for rare point-like mutations caused by nucleotide substitution errors.  Numerical simulations of many successive replications, starting from an arbitrary initial DNA sequence, show that the fractions of mono- and oligonucleotides converge toward compliance with Chargaff's second parity rule.  The theoretical framework developed for this multireplication process demonstrates that the near-equalities of complementary nucleotide fractions arise from two key features: (1)~the dominant role of base-pair complementarity in replication kinetics and (2)~the low intrinsic error rate of DNA polymerases.  Together, these two features yield a robust mechanistic basis for Chargaff's second parity rule.  These considerations explain the existence of deviations with respect to the predictions of models assuming no-strand-bias conditions.
\end{abstract}

\noindent 
\vskip 0.5 cm

\maketitle

\section{Introduction}

The genome of organisms is encoded in deoxyribonucleic acids (DNA) forming copolymeric sequences of nucleotides linked together by phosphodiester bonds \cite{ABLRRW89}.  The four types of nucleotides contain the bases adenine, cytosine, guanine, and thymine, which are more shortly denoted A, C, G, and T, respectively.  Discovered in 1950, Chargaff's first parity rule states that the fractions of complementary nucleotides are equal in DNA, $A=T$ and $C=G$ \cite{C50,C51,CLGH51,CLG52}.  This rule is the direct consequence of complementary base pairing, A:T and C:G, by hydrogen bonds between the two strands of DNA.  The equalities $A=T$ and $C=G$ of Chargaff's first parity rule are thus explained by the interstrand symmetry due to the double helix structure of DNA, established by Watson and Crick in 1953 \cite{WC53a,WC53b}.

Discovered in 1968, Chargaff's second parity rule stipulates the approximate equalities $A\simeq T$ and $C\simeq G$ between the fractions of complementary nucleotides in each one of the two strands of DNA \cite{KRC68II,KRC68III}.  Furthermore, it was discovered that Chargaff's second parity rule extends to the fractions of complementary oligonucleotides in each single strand of DNA \cite{P93,QC01,BHB02,PNN14}.  Today, Chargaff's second parity rule is observed in most genomes: those of bacteria, archaea, eukaryota, and double-stranded DNA viruses.  Deviations with respect to this rule exist for the genomes of mammalian mitochondria and single-stranded DNA or RNA viruses \cite{MB06,NA06}.

Many hypotheses have been formulated in order to understand the origins of Chargaff's second parity rule.  The issue is that this approximate intrastrand symmetry cannot be explained by the double helix structure of DNA, as for Chargaff's first parity rule.  Nevertheless, it is understood that the composition of each one of the two DNA strands is the result of an evolutionary process spanning over many replications, during which mutations have occurred.  In this respect, mathematical models have been proposed for the evolution of mononucleotide fractions under the mutational pressure of nucleotide substitutions.  In particular, under the hypothesis of no-strand-bias conditions for the substitution rates, there is convergence toward strict equalities between the fractions of complementary nucleotides \cite{S95,L95,LL99}.  This model can be generalized to allow a stochastic evolution toward approximately equal fractions with suitable fitting to experimental data \cite{HMO12,FTPM21,PS23}.  If such models can describe the evolution of nucleotide fractions, they do not provide an understanding for the origins of the phenomenon.

Furthermore, several mechanisms have been considered to explain Chargaff's second parity rule. In particular, it has been assumed that the presence of complementary oligonucleotides in DNA sequences could provide an evolutionary advantage by increasing the potential to form stem-loops \cite{F95,BF99a,BF99b}.  Mechanisms have also been considered that are based on mutations by inversions and inverted transpositions \cite{A06,A07} or by inversions only \cite{OWS07}.  In the last two mechanisms, the fractions of complementary nucleotides converge during evolution toward values compliant with Chargaff's second parity rule, while preserving their mean initial values, i.e., $\lim_{r\to\infty} A_r=\lim_{r\to\infty} T_r=(A_0+T_0)/2$ and $\lim_{r\to\infty} C_r=\lim_{r\to\infty} G_r=(C_0+G_0)/2$, featuring a role to primordial genome, although these mean values are observed to be specific to each species.  The causes of Chargaff's second parity rule thus remain elusive.

Here, our purpose is to revisit studies of successive DNA replications based on detailed kinetic models for DNA polymerases, in which numerical evidence was obtained for the maintain of approximate equalities between the fractions of complementary nucleotides \cite{G16PTRSA,G17PRE}.  These numerical simulations gave a hint that many successive DNA replications may lead to compliance with Chargaff's second parity rule.  However, the sole polymerase used in Refs.~\cite{G16PTRSA,G17PRE} is the human mitochondrial DNA polymerase $\gamma$, which is not representative of typical polymerases acting for the replication (or the repair) of the main genome of organisms.  Therefore, we here consider more typical DNA polymerases, namely, the polymerases Dpo1 \cite{ZBNS09}, Dpo3 \cite{BBT12}, and Dpo4 \cite{FS04} from the archaeon {\it Sulfolobus solfataricus P2}, the D-family DNA polymerase from {\it Thermococcus} archaeon species 9$^\circ$N \cite{SG15}, and the wild-type rat DNA polymerase~$\beta$ \cite{AWT97}.   The polymerization rate constants and the dissociation constants of these polymerases have been measured experimentally with pre-steady-state kinetic methods.  The knowledge of these constants can be used for the numerical simulation of many successive replications, starting from arbitrary initial templates.  At each replication, the simulation proceeds by stochastic events generated with Gillespie's kinetic Monte Carlo algorithm \cite{G76,G77}, attaching or detaching a nucleotide to the end of the copy growing along a template.  This process is repeated many successive times.  The fractions of mono- and oligonucleotides are numerically measured in each copy in order to determine their evolution with the number of replications and to test the validity of Chargaff's second parity rule.

After this introduction, the paper is organized as follows.  In Sections~\ref{sec:simul-B} and~\ref{sec:math-theory}, we consider the case where the kinetics of DNA polymerases do not depend on the previously incorporated nucleotide.  Section~\ref{sec:simul-B} presents the results of numerical simulations for the five DNA polymerases here considered. Section~\ref{sec:math-theory} is devoted to the theory of the simulations, showing that the nucleotide fractions converge toward asymptotic values complying with Chargaff's second parity rule in the sense that, asymptotically, the fractions of complementary nucleotides are approximately equal to each other up to unequal corrections of the order of the polymerase error probability.  In Section~\ref{sec:Markov}, the study is extended to polymerase kinetics with a dependence on the previously incorporated nucleotide, confirming the convergence toward compliance with Chargaff's second parity rule.  Conclusion and perspectives are drawn in Section~\ref{sec:Conclusion}.

\section{Simulation of many successive DNA replications}
\label{sec:simul-B}

\subsection{Model of successive DNA replication}

\begin{figure}[h]
\includegraphics[width=10.5cm]{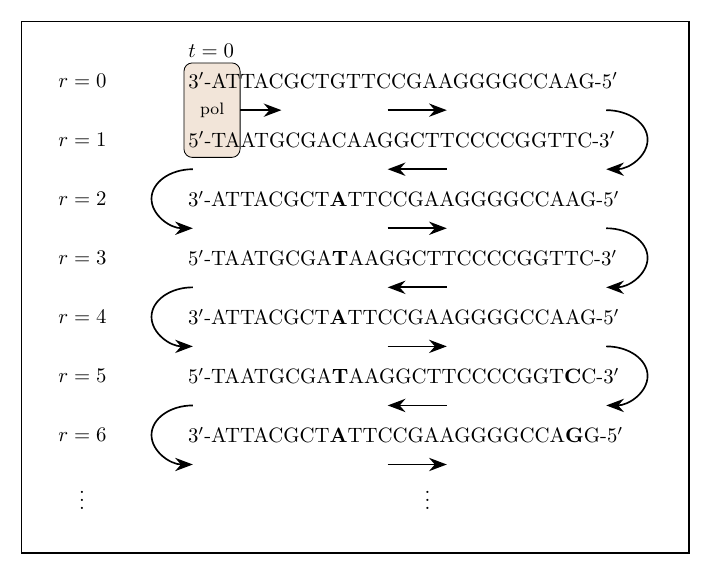}
\caption{Schematic representation of many successive DNA replications $r=1,2,3,\dots$. Some DNA polymerase (pol) is depicted at time $t=0$ at the beginning of the replication of the template $r=0$ into its copy $r=1$, which is further replicated into the copies  $r=2,3,\dots$.  Every replication proceeds from the $3'$ to the $5'$ chain end of the template, hence the zigzag progress of the successive replications.  Replication errors may happen substituting a few nucleotides into non-complementary ones (in bold).  The successive replications may be achieved by several DNA polymerases of the same type.}
\label{fig1}
\end{figure}

We consider a model where many DNA replications are successively performed by some type of polymerase.  Every replication is a template-directed copolymerization process generating a new DNA strand from the previous one, as schematically represented in Fig.~\ref{fig1}.  Every DNA strand is a chain of monomeric units, which are the deoxyribonucleotides A, C, G, or T.  The polymerase always moves from the $3'$ chain end of the template toward its $5'$ chain end.  The copy is synthesized from its $5'$ to its $3'$ chain end.  

Every replication of a template sequence into a copy sequence is generated by random events attaching (and possibly detaching) nucleotides to the copy under the catalytic action of the polymerase according to the following reaction scheme:
\be
 \begin{array}{l}
 \end{array}
\begin{array}{l}
\quad\ \ \, {\rm copy}  = m_1m_2\cdots m_{l-1} \qquad\qquad\quad + \ m_{l}{\rm PP}\\
{\rm template}  = n_1\ n_2\; \cdots \, n_{l-1}\, n_{l} \ n_{l+1}\ \cdots
\end{array}
  \quad \xrightleftharpoons[W_{-m_l,l}]{W_{+m_l,l}}
  \quad
  \begin{array}{l}
\quad\ \ \, {\rm copy}'=m_1m_2\cdots m_{l-1} m_{l} \qquad\qquad + \ {\rm PP}\\
{\rm template}\; =n_1\ n_2\; \cdots \, n_{l-1}\, n_{l} \ n_{l+1}\ \cdots
\end{array}
\quad
\label{reaction}
\ee
where $m_j\in\{{\rm A},{\rm C},{\rm G},{\rm T}\}$ denotes the $j^{\rm th}$ nucleotide in the copy and $n_j\in\{{\rm A},{\rm C},{\rm G},{\rm T}\}$ the $j^{\rm th}$ nucleotide in the template.  Prior to its incorporation as a deoxyribonucleoside monophosphate into the copy, the unit $m_l$ was in the intracellular aqueous solution in the form $m_l{\rm PP}$ of a deoxyribonucleoside triphosphate dATP, dCTP, dGTP, or dTTP.  Upon its binding to the copy, a pyrophosphate PP is released into the solution.

The attachment and detachment rates are denoted $W_{\pm m_l,l}$, respectively.  They depend on the intracellular concentrations of dATP, dCTP, dGTP, and dTTP at the time of the replication.  The copy grows if the nucleotide concentrations are high enough for the attachment rates $W_{+m_l,l}$ to be sufficiently larger than the detachment rates $W_{-m_l,l}$.  As for many other enzymes \cite{MM13,S81,L82}, the dependence of the rates on the concentrations is determined by the Michaelis-Menten kinetics of DNA polymerases, which basically proceeds in two steps. The first step is the rapid binding and unbinding of the nucleoside triphosphate $m_l{\rm PP}$ to the template unit $n_l$ by hydrogen bonds, which step is at quasi-equilibrium with Michaelis-Menten constant $K_{m_l\atop n_l}$.  The second step is the slower formation of the phosphodiester bond itself with the polymerization rate constant $k^{\rm p}_{+m_l\atop \ \, n_l}$.  These constants characterize the kinetics of each type of polymerases and they can be measured experimentally \cite{J93}.  Table~\ref{tab:pol-param} gives the values of these constants for several types of polymerases.

\begin{table}[h]
\caption{Kinetic parameters for the exonuclease-deficient DNA polymerases of this study: the DNA polymerases Dpo1 \cite{ZBNS09}, Dpo3 \cite{BBT12}, and Dpo4 \cite{FS04} from {\it Sulfolobus solfataricus P2}, the D-family DNA polymerase from {\it Thermococcus} archaeon species~9$^\circ$N \cite{SG15}, and the wild-type rat DNA polymerase $\beta$ \cite{AWT97}.  These parameters have been measured experimentally using pre-steady-state kinetic methods. $k^{\rm p}_{+m\atop \ \, n}$ denotes the polymerization rate constant for the incorporation of the copy nucleotide~$m$ paired with the template nucleotide~$n$ and $K_{m\atop n}$ is the dissociation constant of the pair $m\!:\!n$.}
\label{tab:pol-param}
\vspace{5mm}
\begin{center}
\begin{tabular}{|c|cc|cc|cc|cc|cc|}
\hline
&\multicolumn{2}{c|}{{\it Sulf. solf.} Dpo1} &\multicolumn{2}{c|}{{\it Sulf. solf.} Dpo3} &\multicolumn{2}{c|}{{\it Sulf. solf.} Dpo4} &\multicolumn{2}{c|}{{\it Therm.} sp. polD} &\multicolumn{2}{c|}{WT rat pol $\beta$} \\
\hline
$m$:$n$ & \ $k^{\rm p}_{+m\atop \ \, n}$ & $K_{m\atop n}$ & $k^{\rm p}_{+m\atop \ \, n}$ & $K_{m\atop n}$ & \ $k^{\rm p}_{+m\atop \ \, n}$ & $K_{m\atop n}$& \ $k^{\rm p}_{+m\atop \ \, n}$ & $K_{m\atop n}$& \ $k^{\rm p}_{+m\atop \ \, n}$ & $K_{m\atop n}$ \\
pair & \ (s$^{-1}$) & ($\mu$M) &  (s$^{-1}$) & ($\mu$M) & \ (s$^{-1}$) & ($\mu$M) & \ (s$^{-1}$) & ($\mu$M) &  \ (s$^{-1}$) & ($\mu$M) \\
\hline
A:T & $11.5$ & $4.9$ & $0.12$ & $18$  & $16.1$ & $206$  & $2.6$ & $2.5$  & $24.1$ & $52$ \\
A:G & $0.7$ & $2633$& $0.0025$ & $1900$  & $0.009$ & $334$  & $0.12$ & $320$  & $0.0162$ & $210$ \\
A:C & $0.34$ & $2300$& $0.00065$ & $740$  & $0.016$ & $617$  & $1.3$ & $570$  & $ 0.025$ & $890$ \\
A:A & $1.3$ & $2800$& $0.0010$ & $230$  & $0.006$ & $578$  & $0.15$ & $390$  & $0.006$ & $290$ \\
C:T & $0.7$ & $3300$& $0.0009$ & $790$  & $0.026$ & $309$  & $0.35$ & $530$  & $0.18$ & $630$ \\
C:G & $5.7$ & $8.2$& $0.045$ & $61$  & $7.6$ & $70$  & $3.1$ & $1.7$  & $9.4$ & $8.6$ \\
C:C & $0.009$ & $2000$& $0.0015$ & $813.27$  & $0.034$ & $192$  & $0.42$ & $390$  & $0.012$ & $580$ \\
C:A & $0.7$ & $2633$& $0.0016$ & $2000$  & $0.013$ & $1036$  & $0.22$ & $450$  & $0.020$ & $200$ \\
G:T & $0.9$ & $3200$& $0.0025$ & $410$  & $0.066$ & $935$  & $0.11$ & $300$  & $0.13$ & $850$ \\
G:G & $0.032$ & $900$& $0.0021$ & $690$  & $0.008$ & $131$  & $0.07$ & $320$  & $0.0073$ & $230$ \\
G:C & $5.3$ & $4.1$& $0.069$ & $24$  & $9.4$ & $171$  & $2.1$ & $0.9$  & $13.5$ & $108$ \\
G:A & $0.05$ & $1200$& $0.0017$ & $1700$  & $0.007$ & $1150$  & $0.18$ & $330$  & $0.0079$ & $980$ \\
T:T & $1.7$ & $4500$& $ 0.0015$ & $390$  & $0.034$ & $1941$  & $0.28$ & $490$  & $0.0085$ & $820$ \\
T:G & $1.3$ & $3500$& $0.0016$ & $300$  & $0.077$ & $1283$  & $0.38$ & $650$  & $0.11$ & $380$ \\
T:C & $0.7$ & $2633$& $0.0013$ & $430$  & $0.011$ & $913$  & $0.55$ & $1400$  & $0.018$ & $2000$ \\
T:A & $8.2$ & $11$& $0.038$ & $57$  & $9.4$ & $230$  & $1.8$ & $1.6$  & $16.7$ & $41$ \\
\hline
\end{tabular}
\end{center}
\end{table} 

The fact is that polymerases may catalyze not only the formation of the four Watson-Crick complementary base pairs, but also that of twelve non-complementary base pairs, leading to point-like nucleotide substitutions in the genome.  The attachment rates of the incorrect pairs are significantly smaller than for the correct pairs.  Therefore, polymerases generate replication errors with a rate, which defines the error probability $\eta$ of the polymerase.  For the purpose of the present demonstration, we here consider polymerases, which are devoid of exonuclease proofreading activity, so that their error probability is in the range $\eta\sim 10^{-4}$-$10^{-3}$.  Similar considerations apply to DNA polymerases having their exonuclease activity, which may decrease their error probability, as discussed here below in Subsection~\ref{subsec:exo+pol}.
 
The attachment and detachment rates $W_{\pm m_l,l}$ depend not only on the nucleotide $m_l$ that is incorporated, but also on the other molecular groups constituting its environment upon its binding to the copy.  These molecular groups include the template nucleotide $n_l$, with which the base pair $m_l\!:\!n_l$ is formed, the neighboring template nucleotides $n_{l+1}$ and $n_{l-1}$, and possibly also the nucleotide $m_{l-1}$, which has been previously incorporated into the copy.  In the present Section~\ref{sec:simul-B}, we consider the simplest model where the rates do not depend on the previously formed base pair $m_{l-1}\!:\!n_{l-1}$.  The analytic expressions of the rates of this model are given in Appendix~\ref{app:simul-B}.  In this case, the sequence of the growing copy forms a Bernoulli chain \cite{G16PTRSA,G16PRL,G17JSM}.  The more complicated case, where the rates also depend on the previously formed base pair $m_{l-1}\!:\!n_{l-1}$, will be considered in Section~\ref{sec:Markov}.

The replication of a template into its copy can be numerically simulated with the Gillespie algorithm \cite{G76,G77}.  Such a replication can be repeated many successive times.  Each replication starts again from the $3'$ chain end of the DNA strand, as shown in Fig.~\ref{fig1}.  In every successive copy, the fractions of mononucleotides and oligonucleotides are numerically evaluated in order to plot these fractions as functions of the replication number $r$.  These numerical simulations have been carried out for the five polymerases of Table~\ref{tab:pol-param}.  The error probability $\eta_r$ of the $r^{\rm th}$ replication is defined as the fraction of non-complementary (i.e., incorrect) base pairs between the sequences of the copy and the template in the replication.  Its asymptotic value $\eta$ is obtained in the limit $r\to\infty$.

The numerical simulations of successive replications have been carried out for the three following sets of nucleotide concentrations:
\bea
{\rm concentrations\; I:}&& \quad [{\rm dATP}] = 3.2 \, \mu{\rm M} \, , 
\quad [{\rm dCTP}] = 2.1 \, \mu{\rm M} \, , 
\quad [{\rm dGTP}] = 1.5 \, \mu{\rm M} \, , 
\quad [{\rm dTTP}] = 5.4 \, \mu{\rm M} \, ; \label{conc-I}\\
{\rm concentrations\; II:}&& \quad [{\rm dATP}] = 24 \, \mu{\rm M} \, , 
\quad\; [{\rm dCTP}] = 29 \, \mu{\rm M} \, , 
\quad\; [{\rm dGTP}] = 5.2 \, \mu{\rm M} \, , 
\quad [{\rm dTTP}] = 37 \, \mu{\rm M} \, ; \label{conc-II}\\
{\rm concentrations\; III:}&& \quad [{\rm dATP}] =  [{\rm dCTP}] =  [{\rm dGTP}] =  [{\rm dTTP}] = 100 \, \mu{\rm M} \, .\label{conc-III}
\eea
The concentrations I are the mean physiological nucleotide concentrations in normal resting cells and the concentrations II are those in dividing cells \cite{T94}.  We note that these concentrations are significantly larger than their chemical equilibrium values, which are of the order of nanomolars in the range of $10^{-8}$-$10^{-10}$~M, and where the growth of the copy stops.

\begin{table}[h]
\caption{Asymptotic values of the nucleotide fractions and the corresponding error probability $\eta$ for the numerical simulations with Gillespie's algorithm of many successive DNA replications with the exonuclease-deficient DNA polymerases considered in this study: the DNA polymerases Dpo1, Dpo3, and Dpo4 from {\it Sulfolobus solfataricus P2}, the D-family DNA polymerase from {\it Thermococcus} archaeon species 9$^\circ$N, and the wild-type rat DNA polymerase $\beta$.  In every case, the initial template sequence is random with equal fractions of nucleotides and its length has $L=10^6$ units.  The concentrations I-III are given by Eqs.~(\ref{conc-I})-(\ref{conc-III}).  The statistical errors are smaller than the last digit shown.}
\label{tab:asympt-fractions}
\vspace{5mm}
\begin{center}
\begin{tabular}{|c|c|c|c|c|c|c|}
\hline
DNA polymerase & concentrations & $A$(\%) & $T$(\%) & $C$(\%) & $G$(\%) & $\eta$ \\
\hline
{\it Sulf. solf.} Dpo1 & I & $45.8$ & $45.8$ & $4.2$ & $4.2$ & $0.00054$\\
                         & II & $43.8$ & $43.8$ & $6.2$ & $6.2$ & $0.00066$ \\
                         & III & $33.2$ & $33.2$ & $16.8$ & $16.8$ & $0.00069$ \\
\hline
{\it Sulf. solf.} Dpo3 & I & $47.4$ & $47.3$ & $2.6$ & $2.7$ & $0.0037$\\
                         & II & $46.4$ & $46.2$ & $3.7$ & $3.7$ & $0.0040$ \\
                         & III & $36.7$ & $36.5$ & $13.3$ & $13.5$ & $0.0060$ \\
\hline
{\it Sulf. solf.} Dpo4 & I & $36.6$ & $36.6$ & $13.4$ & $13.4$ & $0.0019$\\
                         & II & $36.1$ & $36.2$ & $14.0$ & $13.7$ & $0.0040$ \\
                         & III & $19.0$ & $19.0$ & $31.0$ & $31.0$ & $0.0022$ \\
\hline
{\it Therm.} sp. polD & I & $40.6$ & $40.6$ & $9.4$ & $9.4$ & $0.0012$\\
                                    & II & $41.7$ & $41.7$ & $8.3$ & $8.3$ & $0.0017$ \\
                                    & III & $23.4$ & $23.4$ & $26.6$ & $26.6$ & $0.0013$ \\
\hline
WT rat pol $\beta$ & I & $34.8$ & $34.8$ & $15.2$ & $15.2$ & $0.00053$\\
                                    & II & $32.9$ & $32.9$ & $17.1$ & $17.1$ & $0.00085$ \\
                                    & III & $17.2$ & $17.2$ & $32.8$ & $32.8$ & $0.00049$ \\
\hline
\end{tabular}
\end{center}
\end{table} 

Simulations have been performed for the following polymerases.

\subsection{{\textbf{\emph{Sulfolobus solfataricus P2}}} DNA polymerase Dpo1}

This B-family polymerase carries out the replication of DNA in the hyperthermophilic archaeon {\it Sulfolobus solfataricus P2} \cite{ZBNS09}.  The genome of this species contains about $3\times 10^6$ base pairs and its nucleotide fractions are $A\simeq T\simeq 32\%$ and $C\simeq G\simeq 18\%$.  The constants of this polymerase have been experimentally measured using pre-steady-state kinetic methods except for three incorrect base pairs, for which typical mean values have been here assumed (see Table~\ref{tab:pol-param}).

The numerical simulation starts from an initial nucleotide template sequence, which is generated as a Bernoulli random chain of length $L=10^6$ with given fractions $A_0$, $C_0$, $G_0$, and $T_0$ for the four nucleotides.  Its replication is numerically simulated using the Gillespie algorithm, which yields a copy given by a new nucleotide sequence.  Next, this new sequence is used as a template for a following replication.  This process is repeated over and over again.  At each replication $r$, the content of the sequence is characterized by the fractions of mononucleotides A, C, G, and~T; and oligonucleotides AT, TA, AC, GT, ATA, TAT, CTA, and TAG; for the purpose of testing Chargaff's second parity rule.  

\begin{figure}[h]
\includegraphics[width=10.5cm]{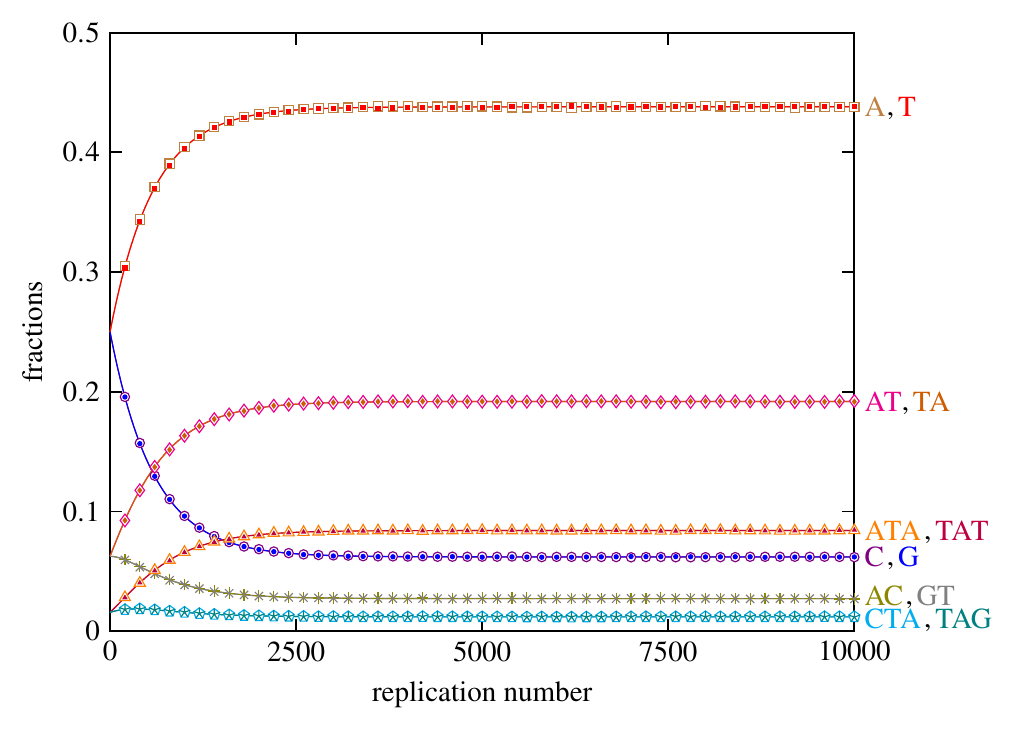}
\caption{Fractions of the mononucleotides A (open squares), T (filled squares), C (open circles), G (filled circles), binucleotides AT (open diamonds), TA (filled diamonds), AC (pluses), GT (times), and trinucleotides ATA (open triangles), TAT (filled triangles), CTA (open pentagons), TAG (stars), versus the even values of the replication number $r$ in DNA strands generated by the DNA polymerase Dpo1 from the archaeon {\it Sulfolobus solfataricus P2} in a solution with the nucleotide concentrations~(\ref{conc-II}).  The initial DNA strand has equal mononucleotide fractions $A_0=C_0=G_0=T_0=25\%$.  The data points showing the results of the numerical simulation with Gillespie's algorithm are plotted every 200 replications.  The solid lines depict the fractions predicted by theory.  The asymptotic values of the fractions are given by $A=T=43.8\%$, $C=G=6.2\%$, $AT=TA=19.2\%$, $AC=GT=2.7\%$, $ATA=TAT=8.4\%$, and $CTA=TAG=1.2\%$, as obtained by averaging their values between the replications $r=9900$ and $r=10000$.}
\label{fig2}
\end{figure}

\begin{figure}[h]
\includegraphics[width=10.5cm]{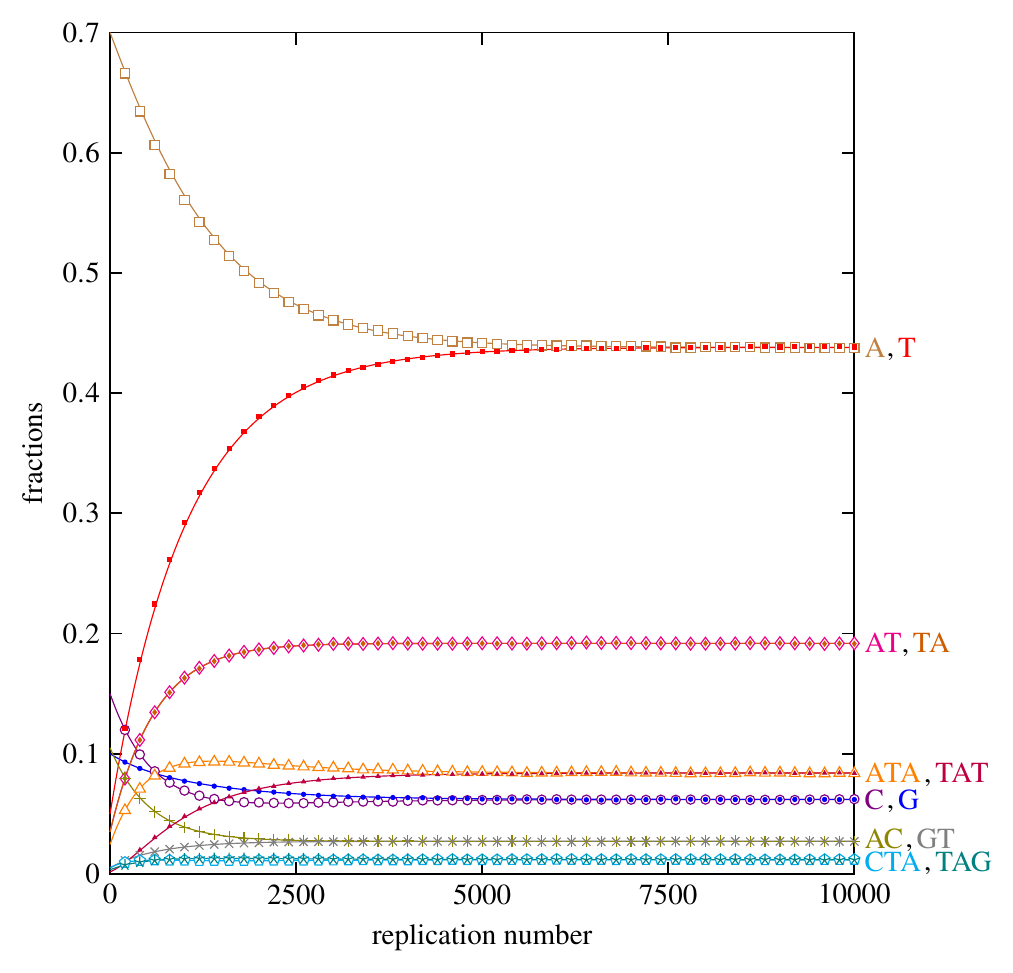}
\caption{Fractions of the mononucleotides A (open squares), T (filled squares), C (open circles), G (filled circles), binucleotides AT (open diamonds), TA (filled diamonds), AC (pluses), GT (times), and trinucleotides ATA (open triangles), TAT (filled triangles), CTA (open pentagons), TAG (stars), versus the even values of the replication number $r$ in DNA strands generated by the DNA polymerase Dpo1 from the archaeon {\it Sulfolobus solfataricus P2} in a solution with the nucleotide concentrations~(\ref{conc-II}).  The initial DNA strand has the unequal mononucleotide fractions $A_0=70\%$, $C_0=15\%$, $G_0=10\%$, and $T_0=5\%$.  The data points showing the results of the numerical simulation with Gillespie's algorithm are plotted every 200 replications.  The solid lines depict the fractions predicted by theory.  The asymptotic values of the fractions are given by $A=T=43.8\%$, $C=G=6.2\%$, $AT=TA=19.2\%$, $AC=GT=2.7\%$, $ATA=TAT=8.4\%$, and $CTA=TAG=1.2\%$, as obtained by averaging their values between the replications $r=9900$ and $r=10000$.}
\label{fig3}
\end{figure}

With the nucleotide concentrations~(\ref{conc-II}), the fractions are shown in Fig.~\ref{fig2} for a simulation starting from an initial template sequence with equal nucleotide fractions $A_0=C_0=G_0=T_0=25\%$ and in Fig.~\ref{fig3} for a simulation starting from an initial sequence with unequal nucleotide fractions $A_0=70\%$, $C_0=15\%$, $G_0=10\%$, and $T_0=5\%$.  Most remarkably, we observe in both cases that the fractions converge toward values that are compliant with Chargaff's second parity rule.  Moreover, the fractions have precisely the same asymptotic values independently of the initial nucleotide sequence, as observed in Figs.~\ref{fig2} and~\ref{fig3}.  These asymptotic values are reported in Table~\ref{tab:asympt-fractions}, together with the corresponding error probability~$\eta$.  The reported values are obtained by averaging over the values between the replications $r=9900$ and $r=10000$.  We see in Fig.~\ref{fig2} that the convergence to the asymptotic values happens after a number of replications of the order of the inverse $\eta^{-1}$ of the error probability (which is known as the fidelity of the polymerase).  However, in Fig.~\ref{fig3}, the convergence to compliant values occurs over a somehow longer number of replications, which will be explained in the next Section~\ref{sec:math-theory}.  Simulations have also been carried out for the two other sets of nucleotide concentrations~(\ref{conc-I}) and~(\ref{conc-III}).  As seen in Table~\ref{tab:asympt-fractions}, the asymptotic values of the four nucleotides are again compliant with Chargaff's second parity rule.  The results also show that the asymptotic values of the nucleotide fractions depend on the intracellular nucleotide concentrations, which is expected since the reaction rates change with the concentrations.

Since the doubling time due to cell division is about 7-8 hr for {\it Sulfolobus solfataricus} \cite{BP97,LMEHB08}, the inverse $\eta^{-1}$ of the polymerase error probability corresponds to a period of $1.2$-$1.7$ yr for converging toward compliance, which is relatively short with respect to evolutionary time scales.

\subsection{{\textbf{\emph{Sulfolobus solfataricus P2}}} DNA polymerase Dpo3}

This other B-family polymerase of the same archaeon is also known to catalyze DNA replication, but at a slower rate and a lower fidelity than Dpo1.  Its kinetic parameters, which have been measured experimentally for the lone polymerase activity \cite{BBT12}, are reported in Table~\ref{tab:pol-param}.

Numerical simulations starting from an initial Bernoulli sequence with equal nucleotide fractions $A_0=C_0=G_0=T_0=25\%$ show that the mononucleotide fractions converge toward asymptotic values over a shorter number of replications than Dpo1, because the error probability $\eta$ is larger for Dpo3.  Accordingly, the asymptotic values can here be calculated by averaging over the replications from $r=900$ to $r=1000$ and they are given in Table~\ref{tab:asympt-fractions}.  These asymptotic values depend on the nucleotide concentrations (\ref{conc-I})-(\ref{conc-III}) considered in the simulations, but they comply with Chargaff's second parity rule.

\subsection{{\textbf{\emph{Sulfolobus solfataricus P2}}} DNA polymerase Dpo4}

This further DNA polymerase from {\it Sulfolobus solfataricus P2} belongs to the family Y in charge of DNA repair.  Its pre-steady-state kinetic parameters have been measured experimentally for the polymerase active site only \cite{FS04} and they are presented in Table~\ref{tab:pol-param}.  Its polymerization rate constants are larger than for Dpo3, but its fidelity is lower than for Dpo1.

For Dpo4, the asymptotic values of the mononucleotide fractions and the error probability are obtained by averaging over the replications between $r=4500$ and $r=5000$.  They are given in Table~\ref{tab:asympt-fractions} for the nucleotide concentrations (\ref{conc-I})-(\ref{conc-III}).  Again, Chargaff's second parity rule is very well satisfied.

\subsection{{\textbf{\emph{Thermococcus}}} species $9^\circ$N DNA polymerase}

{\it Thermococcus} species $9^\circ$N is a thermophilic archaeon with a genome containing about $2\times 10^6$ base pairs.
The polymerase rate constants and dissociation constants have been measured experimentally for its exonuclease-deficient family-D DNA polymerase and their values are reported in Table~\ref{tab:pol-param}.

Numerical simulations of many successive replications again show that the mononucleotide fractions converge toward asymptotic values, which depend on the nucleotide concentrations, but comply with Chargaff's second parity rule.  Table~\ref{tab:asympt-fractions} gives these asymptotic values obtained by averaging over the replications between $r=9900$ and $r=10000$ for the concentrations~(\ref{conc-I})-(\ref{conc-III}).

For genus {\it Thermococcus}, cell division doubling times are observed in the range 24-95 min \cite{RK96,ZZX15}.  In this case, the inverse $\eta^{-1}$ of the polymerase error probability corresponds to a convergence period of 10-55 days, which is very short with respect to evolutionary time scales.

\subsection{Rat DNA polymerase $\beta$}

As a further example, we consider the wild-type (WT) rat DNA polymerase $\beta$.  Rat is a mammal with a genome containing  $2.9\times 10^9$ base pairs in $22$ chromosomes and characterized by the mononucleotide fractions $A\simeq T\simeq 29\%$ and $C\simeq G\simeq 21\%$.  The polymerase $\beta$ belongs to the X-family and acts in DNA repair pathways.  This is an example of small eukaryotic polymerase without any exonuclease activity.  The kinetic parameters of the WT rat DNA polymerase~$\beta$ have been experimentally measured using pre-steady-state kinetic methods and they are also reported in Table~\ref{tab:pol-param} \cite{AWT97}.

For this polymerase, numerical simulations of many successive replications have also been carried out for the concentrations~(\ref{conc-I})-(\ref{conc-III}), starting from an initial Bernoulli sequence with $A_0=C_0=G_0=T_0=25\%$. The convergence of the mononucleotide fractions toward asymptotic values compliant with Chargaff's second parity rule is again observed.  These values are reported in Table~\ref{tab:asympt-fractions} after averaging from $r=9000$ to $r=10000$.

\subsection{Comparison between different polymerases and concentration values}

The results of the numerical simulations show that the asymptotic values of the mononucleotide fractions depend on the type of polymerase carrying out the many successive replications and on the intracellular nucleotide concentrations.  This observation suggests that the nucleotide fractions characterizing the genome of each species have evolved under the combined action of the different polymerases of the species, using intracellular nucleotide concentrations, which may vary in time during the cell cycle between two successive replications.  In this regard, Table~\ref{tab:asympt-fractions} shows that the physiological values (\ref{conc-I}) and~(\ref{conc-II}) of the nucleotide concentrations tend to generate nucleotide fractions $A\simeq T$ that are larger than $C\simeq G$, as often observed in the genome of various organisms \cite{KRC68II,KRC68III,BHB02,OWS07,PNN14}.

Nevertheless, the results of simulations show that Chargaff's second parity rule is always satisfied.  At this stage, this numerical observation remains mysterious because the kinetic parameters reported in Table~\ref{tab:pol-param} take different values for the twelve incorrect base pairs.  The purpose of the following Section~\ref{sec:math-theory} is to explain how Chargaff's second parity rule can hold in spite of such asymmetries.

\section{Theory for successive DNA replications}
\label{sec:math-theory}

\subsection{Theory for a single DNA replication}

During the last decade, important advances have been made in the theory of DNA replication and other template-directed copolymerization processes \cite{GCCR09,C13,AG08,G16a,G16b,G16PRL,G16PTRSA,G17JSM,G17PRE,G20,SSOL17,LZSOL19,LSOL21,QJPBO23,GQO25}.  In particular, exact asymptotic solutions have been obtained for the replication of a template sequence $n_1n_2\cdots n_l\cdots n_L$ into a copy sequence $m_1m_2\cdots m_l\cdots m_L$ for different kinds of kinetics, depending or not on previously incorporated nucleotides, or on conformational changes of the polymerase \cite{G16PRL,G16PTRSA,G17JSM,G17PRE,G20}.  In general, the attachment and detachment rates are specific to the polymerase and they depend on the nucleotide concentrations in the intracellular solution surrounding the polymerase.  For Michaelis-Menten kinetics, the attachment and detachment rates have known mathematical expressions, which are given in Appendix~\ref{app:simul-B}.  For kinetics where the rates are independent of the previous incorporated nucleotide, theory shows that the copy forms a Bernoulli chain, so that the probability of finding some copy sequence grown on a given template sequence factorizes into the product of the probabilities of the individual nucleotides in the copy sequence \cite{G16PRL,G16PTRSA,G17JSM}.

Furthermore, for nucleotide concentrations such as (\ref{conc-I})-(\ref{conc-III}) that are larger than their nanomolar chemical equilibrium values (where detailed balance holds and replication stops), the attachment rates are significantly larger than the detachment rates, so that the latter can be neglected in front of the former, $W_{+m_l,l}\gg W_{-m_l,l}$.  These conditions are satisfied in the numerical simulations of Section~\ref{sec:simul-B}.

Under such circumstances, the conditional probability of finding the nucleotide~$m$ in the copy sequence and paired with the template nucleotide~$n$ can be expressed as
\be
\label{P_mn} 
P_{m\atop n} \equiv \frac{W_{+m\atop \ \, n}}{\sum_{m'}W_{+m'\atop \ \, n}}
\ee
in terms of the attachment rates given by Eq.~(\ref{W+B}).  These conditional probabilities satisfy the following normalization conditions,
\be
\label{P-norm}
\sum_m P_{+m\atop \ \, n}=1
\qquad\mbox{for all} \ n \, .
\ee
If the replication process starts with a template sequence forming a Bernoulli chain, the copy is again a Bernoulli chain, and this property persists upon further replications.

\subsection{Extension to successive DNA replications}

The conditional probabilities~(\ref{P_mn}) allow us to determine the fractions or probabilities of nucleotides in all the copies generated by successive replications according to the process shown in Fig.~\ref{fig1}. The nucleotide probabilities in the $r^{\rm th}$ sequence can be written in the form of the following column matrix or vector:
\be
\label{p_r}
{\bf p}_r = 
\left[
\begin{array}{c}
A_r \\
C_r \\
G_r \\
T_r
\end{array}
\right]
\qquad\mbox{such that}\qquad
A_r+C_r+G_r+T_r=100\% \, .
\ee
Accordingly, the nucleotide probabilities of the sequence generated by the next replication are given by
\be
\label{next-p_r}
{\bf p}_{r+1} = {\boldsymbol{\mathsf P}}\cdot {\bf p}_r 
\ee
in terms of the following $4\times 4$ matrix containing the conditional probabilities~(\ref{P_mn}):
\be
\label{matrix-P}
{\boldsymbol{\mathsf P}} \equiv
\left[
\begin{array}{cccc}
P_{{\rm A}\atop{\rm A}} & P_{{\rm A}\atop{\rm C}} & P_{{\rm A}\atop{\rm G}} & \boxed{ P_{{\rm A}\atop{\rm T}} } \\
P_{{\rm C}\atop{\rm A}} & P_{{\rm C}\atop{\rm C}} & \boxed{ P_{{\rm C}\atop{\rm G}} } & P_{{\rm C}\atop{\rm T}} \\
P_{{\rm G}\atop{\rm A}} & \boxed{ P_{{\rm G}\atop{\rm C}} } & P_{{\rm G}\atop{\rm G}} & P_{{\rm G}\atop{\rm T}} \\
\boxed{ P_{{\rm T}\atop{\rm A}} } & P_{{\rm T}\atop{\rm C}} & P_{{\rm T}\atop{\rm G}} & P_{{\rm T}\atop{\rm T}}
\end{array}
\right] ,
\ee
which defines the matrix of transition probabilities, or more shortly transition matrix, for the process of interest \cite{vK81,S16}.
Probability is conserved, i.e., $A_{r+1}+C_{r+1}+G_{r+1}+T_{r+1}=100\%$ also holds, because of the normalization conditions~(\ref{P-norm}).  The nucleotide probabilities of the $r^{\rm th}$ copy can thus be obtained by $r$ successive matrix multiplications starting from the nucleotide probabilities of the initial template according to ${\bf p}_r = {\boldsymbol{\mathsf P}}^r \cdot {\bf p}_0$.
In particular, the error probability of the $r^{\rm th}$ replication is given by
\be
\eta_r=1-P_{{\rm T}\atop{\rm A}} A_{r-1} -P_{{\rm G}\atop{\rm C}} C_{r-1} - P_{{\rm C}\atop{\rm G}} G_{r-1} -P_{{\rm A}\atop{\rm T}} T_{r-1}
\label{eta_r}
\ee
in terms of the conditional probabilities~(\ref{P_mn}) to form Watson-Crick complementary base pairs and the mononucleotide fractions ${\bf p}_{r-1}$ in the template generating the copy in the replication.  The error probability converges toward its asymptotic value $\eta=\lim_{r\to\infty} \eta_r$, which globally characterizes the replication errors made by the polymerase.

The theoretical predictions based on the matrix~(\ref{matrix-P}) are shown as solid lines in Figs.~\ref{fig2} and~\ref{fig3}.  They agree very well with the results of numerical simulations, confirming the validity of the theoretical assumptions.
We note that the $4\times 4$ matrix~(\ref{matrix-P}) is asymmetric, because polymerase kinetics are specific to the nucleotides involved in each reaction, which are of four different types, so that it is difficult in this regard to understand how the intrastrand symmetries observed in Figs.~\ref{fig2} and~\ref{fig3} can arise.  Yet, the matrix~(\ref{matrix-P}) is not structureless.

The key point is that the (boxed) elements on the anti-diagonal of the matrix~(\ref{matrix-P}) are close to the value of $100\%$, because they give the probabilities of forming Watson-Crick complementary base pairs, while the other elements give the substitution probabilities for the formation of non-complementary base pairs.  Accordingly, the matrix~(\ref{matrix-P}) can be decomposed as
\be
\label{P-decomp}
{\boldsymbol{\mathsf P}} = {\boldsymbol{\mathsf P}}^{(0)} +{\boldsymbol{\mathsf P}}^{(1)}
\ee
into its dominant part,
\be
\label{P0=C}
{\boldsymbol{\mathsf P}}^{(0)} = {\boldsymbol{\mathsf C}} 
\equiv
\left[
\begin{array}{cccc}
0 & 0 & 0 & 1 \\
0 & 0 & 1 & 0 \\
0 & 1 & 0 & 0 \\
1 & 0 & 0 & 0
\end{array}
\right] ,
\ee
and a rest given by another $4\times 4$ matrix ${\boldsymbol{\mathsf P}}^{(1)}$, which is of the order of the error probability $\eta$ of the polymerase and which is thus significantly smaller than its dominant part~(\ref{P0=C}).  The latter is given by an anti-diagonal exchange matrix that directly expresses the Watson-Crick complementarity of the base pairs that are formed most often upon DNA replication.  The rest ${\boldsymbol{\mathsf P}}^{(1)}$ can be considered as a perturbation with respect to the dominant part and treated by perturbation theory using the error probability $\eta$ and related quantities as small perturbation parameters.

\subsection{Asymptotic behavior after many successive replications}

The long-time behavior of the process can be expressed in terms of the four eigenvalues $\lambda$ and the four associated eigenvectors $\bf v$ of the transition matrix~(\ref{matrix-P}), such that ${\boldsymbol{\mathsf P}}\cdot{\bf v}=\lambda\,{\bf v}$.  The leading eigenvalue is equal to $\lambda=1$, because the normalization conditions~(\ref{P-norm}) are always satisfied. Therefore, we can obtain the asymptotic values of the nucleotide fractions or probabilities by calculating the leading eigenvector of the matrix~(\ref{matrix-P}), such that
\be
\label{eq-v}
{\bf v}={\boldsymbol{\mathsf P}}\cdot{\bf v} \, .
\ee
Now, the leading eigenvector can be expanded as the following series,
\be
\label{v-series}
{\bf v}={\bf v}^{(0)} + {\bf v}^{(1)} + {\bf v}^{(2)} + \cdots
\qquad\mbox{with}\qquad
{\bf v}^{(n)} = O(\eta^n) \, ,
\ee
going as powers of the small parameter.  Inserting the decomposition~(\ref{P-decomp}) and the expansion~(\ref{v-series}) into  Eq.~(\ref{eq-v}) and identifying the terms corresponding to the same power of the small parameter, we obtain the following equations at the different orders of perturbation:
\bea
&&\mbox{order 0:}\qquad ({\boldsymbol{\mathsf 1}}-{\boldsymbol{\mathsf C}})\cdot {\bf v}^{(0)} = 0 \, , \label{v-order-0}\\
&&\mbox{order 1:}\qquad ({\boldsymbol{\mathsf 1}}-{\boldsymbol{\mathsf C}})\cdot {\bf v}^{(1)} = {\boldsymbol{\mathsf P}}^{(1)}\cdot{\bf v}^{(0)} \, , \label{v-order-1}\\
&&\qquad\qquad\qquad\qquad\vdots\nonumber
\eea
Given the form~(\ref{P0=C}) of the anti-diagonal exchange matrix~${\boldsymbol{\mathsf C}}$, Eq.~(\ref{v-order-0}) at order~0 gives the following solutions:
\be
\label{v(0)}
{\bf v}^{(0)} = 
\left[
\begin{array}{c}
A^{(0)} \\
C^{(0)} \\
G^{(0)} \\
T^{(0)}
\end{array}
\right]
\qquad\mbox{with}\qquad
A^{(0)}  = T^{(0)}
\qquad\mbox{and}\qquad
C^{(0)} = G^{(0)} \, .
\ee
Therefore, at order~0, we obtain the exact equality between the fractions of complementary base pairs, underlying Chargaff's second parity rule.  However, since this solution is degenerate, we still need to consider Eq.~(\ref{v-order-1}) at order~1, which explicitly reads
\be
\left[
\begin{array}{r}
A^{(1)}-T^{(1)} \\
C^{(1)}-G^{(1)} \\
-C^{(1)}+G^{(1)} \\
-A^{(1)}+T^{(1)} 
\end{array}
\right]
 = {\boldsymbol{\mathsf P}}^{(1)}\cdot
\left[
\begin{array}{c}
A^{(0)} \\
C^{(0)} \\
C^{(0)} \\
A^{(0)}
\end{array}
\right]
 \, ,
\label{v-order-1-bis}
\ee
because of Eqs.~(\ref{P0=C}) and~(\ref{v(0)}).

Multiplying this equation from its left-hand side by the row matrices $[1\, 0\, 0\, 1]$ and $[0\, 1\, 1\, 0]$, we obtain the following expressions for the asymptotic nucleotide fractions at order~0:
\bea
A^{(0)} = T^{(0)} = \frac{P_{{\rm A}\atop{\rm C}} + P_{{\rm A}\atop{\rm G}} + P_{{\rm T}\atop{\rm C}} + P_{{\rm T}\atop{\rm G}} }{2\left(P_{{\rm A}\atop{\rm C}} + P_{{\rm A}\atop{\rm G}} + P_{{\rm T}\atop{\rm C}} + P_{{\rm T}\atop{\rm G}} + P_{{\rm C}\atop{\rm A}} + P_{{\rm G}\atop{\rm A}} + P_{{\rm C}\atop{\rm T}} + P_{{\rm G}\atop{\rm T}}\right)} \, , \label{A0=T0}\\
C^{(0)} = G^{(0)} = \frac{ P_{{\rm C}\atop{\rm A}} + P_{{\rm G}\atop{\rm A}} + P_{{\rm C}\atop{\rm T}} + P_{{\rm G}\atop{\rm T}} }{2\left(P_{{\rm A}\atop{\rm C}} + P_{{\rm A}\atop{\rm G}} + P_{{\rm T}\atop{\rm C}} + P_{{\rm T}\atop{\rm G}} + P_{{\rm C}\atop{\rm A}} + P_{{\rm G}\atop{\rm A}} + P_{{\rm C}\atop{\rm T}} + P_{{\rm G}\atop{\rm T}}\right)} \, . \label{C0=G0}
\eea
Next, if we multiply Eq.~(\ref{v-order-1-bis}) from its left-hand side by the row matrices $[1\, 0\, 0\, 0]$ and $[0\, 1\, 0\, 0]$, we obtain the corrections to the nucleotide fractions at next order: $A^{(1)}-T^{(1)} = O(\eta)\ne 0$ and $C^{(1)}-G^{(1)} = O(\eta)\ne 0$.  These corrections do not satisfy the strict equalities between the fractions of the complementary base pairs, but the differences are of the order of the error probability $\eta$ of the polymerase, which is typically very small.  The intrastrand symmetry is thus only approximate in agreement with observations.  Therefore, this reasoning explains Chargaff's second parity rule stated as the {\it approximate} equality between the intrastrand fractions of complementary nucleotides, since the equality only holds with an accuracy of the order of the polymerase error probability.

We note that, in the particular case where the replication process was homogeneous with conditional probabilities that are all equal to $P_{m\atop n}=1-\eta$ for complementary pairs $m=\tilde n$ and to $P_{m\atop n}=\eta/3$ for non-complementary pairs $m\ne\tilde n$, the asymptotic nucleotide fractions would all be equal to each other with $A=C=G=T=25\%$.

Since the successive copies form Bernoulli chains, the fractions of the oligonucleotides are here given by products of mononucleotide fractions as $p(m_1m_2\cdots m_l)=\prod_{j=1}^l p(m_j)$.  This prediction is indeed confirmed by the results of the numerical simulations shown in Figs.~\ref{fig2} and~\ref{fig3}.

\subsection{Transient behavior before convergence}

The transient behavior of the replication process can be characterized in terms of the whole set of four eigenvalues and eigenvectors  of the matrix~(\ref{matrix-P}).   The leading eigenvalue is always equal to one, but the next-to-leading eigenvalues are not and they can be expanded in powers of the error probability $\eta$ as $\lambda=\lambda^{(0)}+\lambda^{(1)}+\lambda^{(2)}+\cdots$ with $\lambda^{(n)}=O(\eta^n)$.  At order 0, the four eigenvalues $\lambda^{(0)}$ are given by $\{+1,+1,-1,-1\}$.  At order 1, the corrections $\lambda^{(1)}$ are evaluated as $\{0,O(\eta),O(\eta),O(\eta)\}$, respectively. Typically, the three non-leading eigenvalues have an absolute value lower than one, $\vert\lambda\vert < 1$, expressing the fact that the nucleotide fractions converge toward the values fixed by the leading eigenvector associated with the eigenvalue $\lambda=1$.  There are thus three modes of relaxation toward the asymptotic values: a mode of exponential relaxation corresponding to $0<\lambda<1$ and two other modes corresponding to $0<\vert\lambda\vert < 1$ with $\lambda^{(0)}=-1$.  For each mode, the time scale of relaxation is given by $\rho=(-\ln\vert\lambda\vert)^{-1}$, which is measured in replication number.

\subsection{Example}

As an example, we consider the multireplication process for the DNA polymerase Dpo1 from the archeon {\it Sulfolobus solfataricus P2} at the physiological concentrations~(\ref{conc-II}).  In this case, the matrix~(\ref{matrix-P}) can be formed with the elements~(\ref{P_mn}) using the polymerization rate and dissociation constants given in Table~\ref{tab:pol-param} and its eigenvalues $\lambda$ can be computed to be $\{1, 0.998292, -0.999092, -0.998519\}$. The eigenvector associated with the leading eigenvalue $\lambda=1$ and its approximation at order 0, which is given by Eqs.~(\ref{A0=T0}) and~(\ref{C0=G0}), have the following  evaluations, respectively:
\be
\label{v-Sulf.Solf.}
{\bf v} = 
\left[
\begin{array}{r}
43.7985\% \\
6.1930\% \\
6.2039\% \\
43.8046\%
\end{array}
\right]
\qquad\mbox{and}\qquad
{\bf v}^{(0)} = 
\left[
\begin{array}{r}
43.8001\% \\
6.1999\% \\
6.1999\% \\
43.8001\%
\end{array}
\right]
,
\ee
confirming that the equality between the fractions of complementary bases is only approximate, although the differences are of the order of the polymerase error probability $\eta=0.00066$, which is very small, and, thus, undetectable with the accuracy used in Table~\ref{tab:asympt-fractions} for presenting the results of numerical simulations with this polymerase.  Nevertheless, we note that small deviations with respect to the strict equality are visible for the cases where the error probability is the largest in Table~\ref{tab:asympt-fractions}, which supports the statement that approximate equality between complementary-base fractions is controlled by the smallness of the polymerase error probability.

Moreover, going back to the polymerase Dpo1 at concentrations~(\ref{conc-II}), the relaxation time scales of the replication process can be evaluated by $(-\ln\vert\lambda\vert)^{-1}$ in terms of the four eigenvalues $\lambda$ and they are given by $\{\infty, 584.8, 1100.8, 674.9 \}$, respectively.  These values are the numbers of replications, over which the amplitude of the relaxation mode is reduced by a factor ${\rm e}\simeq 2.718$.  For a process starting from equal nucleotide fractions, the approximate equalities $A_r\simeq T_r$ and $C_r\simeq G_r$ are maintained since the beginning of the process, as seen in Fig.~\ref{fig2}.  In this case, the relaxation toward the asymptotic values~(\ref{v-Sulf.Solf.}) is essentially controlled by the eigenvalue $\lambda=0.998292$ corresponding to a relaxation over the time scale $(-\ln\vert\lambda\vert)^{-1}=584.8$, which is observed in Fig.~\ref{fig2}.  However, if the process starts from unequal nucleotide fractions, the relaxation involves all the eigenvalues and, in particular, the eigenvalues with a negative real part, which generate nucleotide fractions that alternate between odd and even replication numbers.   This alternating behavior is not seen in Fig.~\ref{fig3} because the fractions are plotted therein only at even replication numbers.  The fractions at odd replication numbers are actually close to the plotted fractions of the complementary base.  The relaxation mode associated with the eigenvalue $\lambda=-0.999092$ is slower since its relaxation time scale is given by $(-\ln\vert\lambda\vert)^{-1}=1100.8$, which is indeed what is observed in Fig.~\ref{fig3}.  The other eigenvalue $\lambda=-0.998519$ has a similar effect, but over the intermediate time scale $(-\ln\vert\lambda\vert)^{-1}=674.9$.

Therefore, the theory can explain the multireplication process in great detail.

\subsection{Evolution every two replications}

As illustrated here above in Fig.~\ref{fig3}, the evolution of the intrastrand DNA composition is smooth if the nucleotide fractions are plotted versus either even or odd replication numbers, in particular, every two replications.  In this way, the alternating behavior caused by complementarity is no longer apparent.  Furthermore, relaxation toward stationarity has characteristic replication numbers going as the inverse of the error probability, so that the time scale of relaxation is much longer than one or two replications: $\rho=(-\ln\vert\lambda\vert)^{-1}=O(\eta^{-1}) \gg 1$.  Accordingly, every two replications, the nucleotide fractions can be considered as smooth functions ${\bf p}(t)={\bf p}_r$ of the time $t=r\tau$, where $\tau$ is the lapse of time between two replications and $r$ is assumed to remain either an even or an odd integer.  With these assumptions, linear ordinary differential equations, ruling the time evolution of the nucleotide fractions, can be deduced, as shown in Appendix~\ref{app:ODEs}.  

As a consequence of the structure~(\ref{P-decomp})-(\ref{P0=C}) of the transition matrix, the leading terms of these differential equations (which are proportional to the polymerase error probability) obey the no-strand-bias conditions \cite{S95,L95,LL99}, generating the strict intrastrand symmetry.  However, these differential equations have additional terms that are proportional to the square of the polymerase error probability and that break the strict intrastrand symmetry.  Thus, the mechanism described here does not satisfy the no-strand-bias conditions, which confirms that the equalities between the complementary nucleotide fractions are not strict, but only approximate.

\section{Kinetics with a dependence on the previously incorporated nucleotide}
\label{sec:Markov}

In general, the kinetics of polymerases depend not only on the ultimate pair $m_l\!:\!n_l$ of nucleotides, which is formed during the elongation of the copy, but also on the previously incorporated nucleotide $m_{l-1}$ and, thus, on the penultimate pair $m_{l-1}\!:\!n_{l-1}$ \cite{G16PRL,G16PTRSA,G17JSM,G17PRE}.  Such dependences have been measured experimentally.  In particular, polymerases are known to slow down if the previously incorporated nucleotide is incorrect \cite{J93}.  Therefore, it is important to extend the study to such more complicated kinetics in order to evaluate the consequences of the dependence on further nucleotides.

\subsection{Numerical simulations}

As before, the replication process~(\ref{reaction}) can be numerically simulated using Gillespie's algorithm \cite{G76,G77}, now with an extended dependence of the attachment and detachment rates $W_{\pm m_l,l}$ on the penultimate pair $m_{l-1}\!:\!n_{l-1}$.  Assuming that the polymerase has Michaelis-Menten kinetics, the rates are given by expressions similar to Eqs.~(\ref{W+B}) and (\ref{W-B}), but with polymerization rate constants and dissociation constants depending on the previously formed base pair $m_{l-1}\!:\!n_{l-1}$, as explained in Appendix~\ref{app:A3}.  Nevertheless, experimental data are sparse and these constants are often only known if the pairs are either correct (c) or incorrect (i) \cite{J93}.

Here, we focus on the exonuclease-deficient DNA polymerase Dpo1 from the archaeon {\it Sulfolobus solfataricus P2}, taking the  values of Table~\ref{tab:pol-param} for the polymerization rate constants $k^{\rm p}_{{+m_{l}\atop \ \, n_{l}}\!\big\vert{\rm c}}=k^{\rm p}_{+m_{l}\atop \ \, n_{l}}$ and the dissociation constants $K_{{m_{l}\atop n_{l}}\!\big\vert{\rm c}}=K_{m_{l}\atop n_{l}}$ after correct incorporation; and the polymerization rate constants after incorrect incorporation as
\be
k^{\rm p}_{+{\rm c}\vert{\rm i}} = 0.13\; {\rm s}^{-1}\, , \qquad
k^{\rm p}_{+{\rm i}\vert{\rm i}} = 0.001\; {\rm s}^{-1}\, , 
\label{kp-cc-ci-M}
\ee
and the corresponding dissociation constants as
\be
K_{{\rm c}\vert{\rm i}} = 1000 \; \mu{\rm M} \, , \qquad  
K_{{\rm i}\vert{\rm i}} = 2600 \; \mu{\rm M} \, ,
\label{K-cc-ci-M}
\ee
for ${{m_{l}\atop n_{l}}\!\big\vert\!{m_{l-1}\atop n_{l-1}}}={{\rm c}\vert{\rm i}} $ or ${{\rm i}\vert{\rm i}}$, respectively \cite{ZBNS09}.

\begin{figure}[h]
\includegraphics[width=10.5cm]{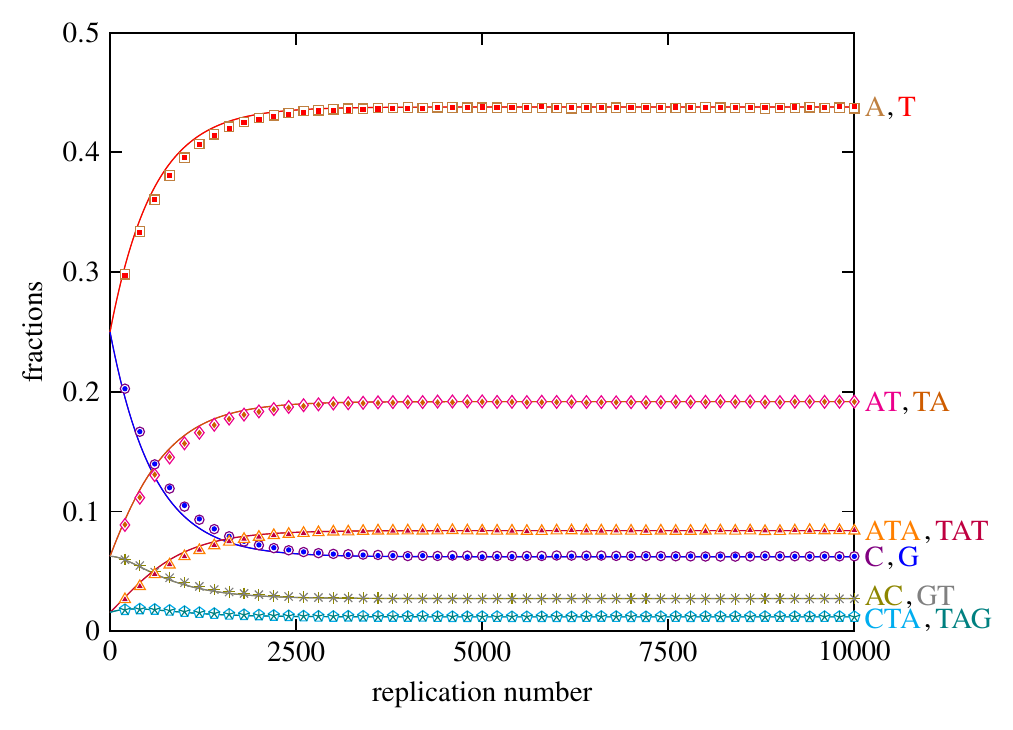}
\caption{Fractions of the mononucleotides A (open squares), T (filled squares), C (open circles), G (filled circles), binucleotides AT (open diamonds), TA (filled diamonds), AC (pluses), GT (times), and trinucleotides ATA (open triangles), TAT (filled triangles), CTA (open pentagons), TAG (stars), versus the even values of the replication number $r$ in DNA strands generated by the DNA polymerase Dpo1 from the archaeon {\it Sulfolobus solfataricus P2} in a solution with the nucleotide concentrations~(\ref{conc-II}), here, for the kinetics depending on the previously incorporated nucleotide with the constants~(\ref{kp-cc-ci-M}) and~(\ref{K-cc-ci-M}).  The initial DNA strand has equal mononucleotide fractions $A_0=C_0=G_0=T_0=25\%$.  The data points showing the results of the numerical simulation with Gillespie's algorithm are plotted every 200 replications.  The solid lines depict the fractions predicted by theory.  The asymptotic values of the fractions are given by $A=T=43.8\%$, $C=G=6.2\%$, $AT=TA=19.2\%$, $AC=GT=2.7\%$, $ATA=TAT=8.4\%$, and $CTA=TAG=1.2\%$, as obtained by averaging their values between the replications $r=9900$ and $r=10000$.}
\label{fig4}
\end{figure}

\begin{figure}[h]
\includegraphics[width=10.5cm]{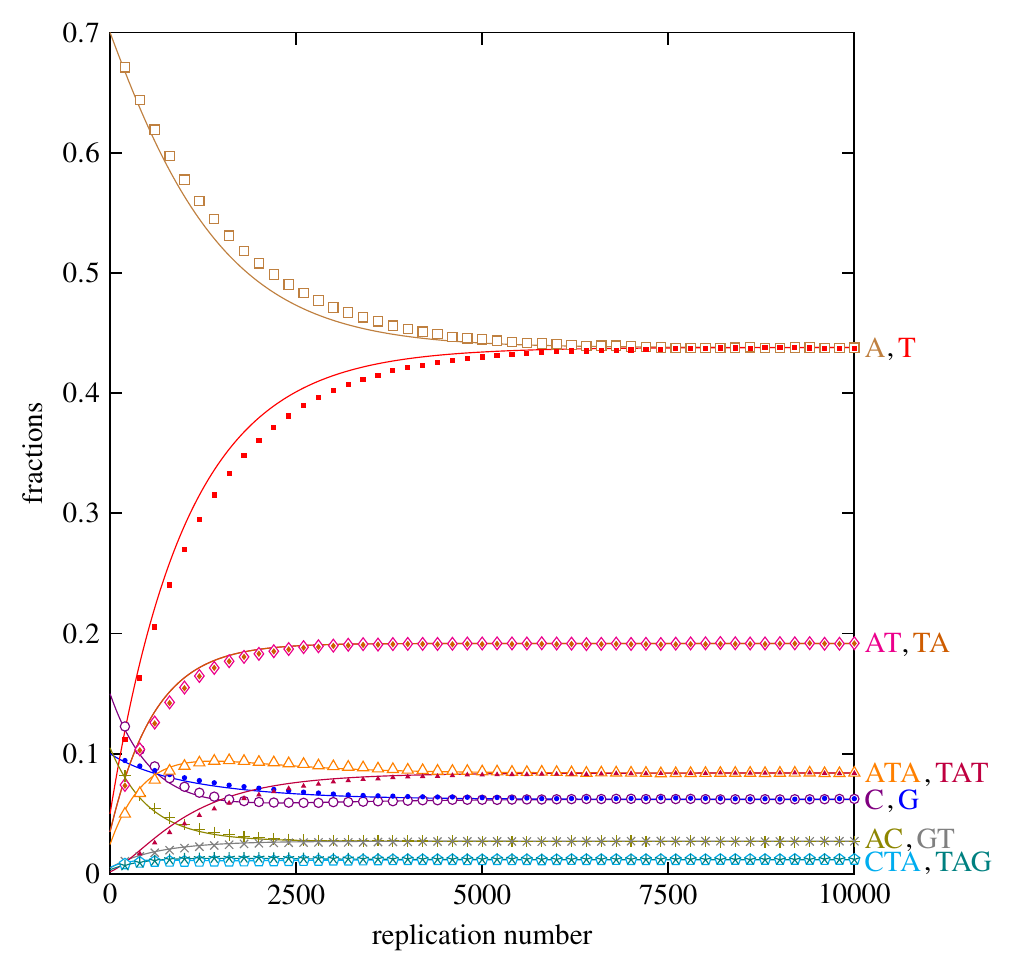}
\caption{Fractions of the mononucleotides A (open squares), T (filled squares), C (open circles), G (filled circles), binucleotides AT (open diamonds), TA (filled diamonds), AC (pluses), GT (times), and trinucleotides ATA (open triangles), TAT (filled triangles), CTA (open pentagons), TAG (stars), versus the even values of the replication number $r$ in DNA strands generated by the DNA polymerase Dpo1 from the archaeon {\it Sulfolobus solfataricus P2} in a solution with the nucleotide concentrations~(\ref{conc-II}), here, for the kinetics depending on the previously incorporated nucleotide with the constants~(\ref{kp-cc-ci-M}) and~(\ref{K-cc-ci-M}).  The initial DNA strand has the unequal mononucleotide fractions $A_0=70\%$, $C_0=15\%$, $G_0=10\%$, and $T_0=5\%$.  The data points showing the results of the numerical simulation with Gillespie's algorithm are plotted every 200 replications.  The solid lines depict the fractions predicted by theory.  The asymptotic values of the fractions are given by $A=T=43.8\%$, $C=G=6.2\%$, $AT=TA=19.2\%$, $AC=GT=2.7\%$, $ATA=TAT=8.4\%$, and $CTA=TAG=1.2\%$, as obtained by averaging their values between the replications $r=9900$ and $r=10000$.}
\label{fig5}
\end{figure}

The results of the numerical simulations are shown in Fig.~\ref{fig4} starting from an initial template with equal nucleotide fractions (as in Fig.~\ref{fig2}); and in Fig.~\ref{fig5} starting from initial unequal fractions (as in Fig.~\ref{fig3}); in both cases with the nucleotide concentrations~(\ref{conc-II}). Again, after many successive replications, we observe the convergence of the mono- and oligonucleotide fractions toward asymptotic values complying with Chargaff's second parity rule.  Numerical simulations have also been performed for the two other sets of concentrations~(\ref{conc-I}) and~(\ref{conc-III}) and the asymptotic mononucleotide fractions of these simulations are presented in Table~\ref{tab:asympt-fractions-M}.

\begin{table}[h]
\caption{Asymptotic values of the nucleotide fractions and the corresponding error probability $\eta$ for the numerical simulations with Gillespie's algorithm of many successive DNA replications with the exonuclease-deficient DNA polymerase from {\it Sulfolobus solfataricus P2}, here, for the kinetics depending on the previously incorporated nucleotide with the constants~(\ref{kp-cc-ci-M}) and~(\ref{K-cc-ci-M}). In every case, the initial template sequence is random with equal fractions of nucleotides and its length has $L=10^6$ units.  The concentrations I-III are given by Eqs.~(\ref{conc-I})-(\ref{conc-III}).  The statistical errors are smaller than the last digit shown.}
\label{tab:asympt-fractions-M}
\vspace{5mm}
\begin{center}
\begin{tabular}{|c|c|c|c|c|c|c|}
\hline
DNA polymerase & concentrations & $A$(\%) & $T$(\%) & $C$(\%) & $G$(\%) & $\eta$ \\
\hline
{\it Sulf. solf.} Dpo1 & I & $45.3$ & $45.3$ & $4.7$ & $4.7$ & $0.00027$\\
                         & II & $43.8$ & $43.8$ & $6.2$ & $6.2$ & $0.00057$ \\
                         & III & $33.0$ & $33.0$ & $17.0$ & $17.0$ & $0.00066$ \\
\hline
\end{tabular}
\end{center}
\end{table} 

In Figs.~\ref{fig4} and~\ref{fig5}, we also observe that the inclusion of a dependence on the previously incorporated nucleotide does not significantly change the results.  The asymptotic values of the nucleotide fractions are modified, but by small amounts, as seen by comparing the values given in Tables~\ref{tab:asympt-fractions} and~\ref{tab:asympt-fractions-M} for the polymerase Dpo1 from {\it Sulfolobus solfataricus P2}.  The results generated by the kinetics with a dependence on the previously incorporated nucleotide turn out to be close to those of Section~\ref{sec:simul-B}, which have been obtained without this dependence.

\subsection{Theory}

Theory shows that, in general, kinetics depending on penultimate pairs generate sequences forming Markov chains \cite{G16PRL,G16PTRSA,G17JSM,G17PRE}.  As aforementioned, for the concentration sets~(\ref{conc-I})-(\ref{conc-III}), the detachment rates are negligible in front of the attachment rates, $W_{+m_l,l}\gg W_{-m_l,l}$.

Under such circumstances, the conditional probability to form the ultimate pair $m\!:\!n$ provided that the penultimate pair (x) is either correct (c) or incorrect (i) is given by
\be
\label{P_mn-M} 
P_{{m\atop n}\!\big\vert{\rm x}} \equiv \frac{W_{{+m\atop \ \, n}\!\big\vert{\rm x}}}{\sum_{m'}W_{{+m'\atop \ \, n}\!\big\vert{\rm x}}}
\qquad\mbox{for}\qquad
\mbox{x = c, i}
\ee
in terms of the corresponding attachment rates $W_{{+m\atop \ \, n}\!\big\vert{\rm x}}$ \cite{G16PRL,G17PRE}.

After many successive replications, the error probability of the replication process reaches the asymptotic value $\eta$ given in Table~\ref{tab:asympt-fractions-M}.  Therefore, along a template sequence, the fraction of correct penultimate pairs is equal to $(1-\eta)$ and the fraction of incorrect ones to $\eta$.  Thus, a possible approximation consists in supposing that the matrix of conditional probabilities can be taken as an average over correct and incorrect penultimate pairs as
\be
\label{matrix-P-M}
{\boldsymbol{\mathsf P}} = (1-\eta)\, {\boldsymbol{\mathsf P}}_{\rm c} + \eta\, {\boldsymbol{\mathsf P}}_{\rm i}
\qquad\mbox{with}\qquad
{\boldsymbol{\mathsf P}}_{\rm x} = \Big[ P_{{m\atop n}\!\big\vert{\rm x}} \Big]
\quad\mbox{for}\quad
\mbox{x = c, i} \, .
\ee
Therefore, the nucleotide fractions or probabilities after $r$ replications can be evaluated by matrix multiplication according to ${\bf p}_r = {\boldsymbol{\mathsf P}}^r \cdot {\bf p}_0$, where ${\bf p}_0$ is the vector of initial fractions.  In this approximation, the memory of the previously incorporated nucleotide is neglected and the effective replication process reduces to one generating Bernoulli chains (rather than Markov chains).  

In Figs.~\ref{fig4} and~\ref{fig5}, the solid lines show the results of these theoretical predictions.  We observe agreement between the approximations based on the matrix~(\ref{matrix-P-M}) and the results of the numerical simulation of the complete process with Gillespie's algorithm.  The agreement is excellent for the asymptotic values of the mono- and oligonucleotide fractions.  This observation shows that the sequences generated by the replication process are close to be Bernoulli chains and that they have a negligible Markovian character.  Nevertheless, deviations between theory and numerical simulations are observed in the transient behavior.  A possible reason for these deviations is that the error probability $\eta_r$ depends in general on the sequence that is copied and, thus, on the replication number $r$.  Therefore, the approximation~(\ref{matrix-P-M}), which consists in replacing the current error probability by its asymptotic value $\eta=\lim_{r\to\infty}\eta_r$, can only be valid in the asymptotic regime and should not be expected to apply during transients.

\subsection{DNA polymerases with exonuclease proofreading}
\label{subsec:exo+pol}

In addition to the domain of their genuine polymerase activity, most DNA polymerases have another domain with an exonuclease activity, which has a proofreading effect on replication \cite{ABLRRW89}.  This error-correction mechanism uses the dependence of the polymerase activity on the previously incorporated nucleotide \cite{J93}.  If polymerization is slowed down by the insertion of an incorrect nucleotide, the DNA chain has the time to jump to the exonuclease domain, where the incorrect nucleotide is cleaved by hydrolysis, lowering the error probability $\eta$ by a factor of about $10^2$.  Kinetic theory shows that the rates of exonuclease activity are added to the rates of the polymerase activity \cite{G16b}. Since the error probability is lower, complementarity is stronger between the copy and the template upon replication, reinforcing the dominant role of the anti-diagonal exchange matrix~(\ref{P0=C}) at the basis of Chargaff's second parity rule.  However, the convergence toward compliance would be delayed by the factor enhancing the fidelity of such DNA polymerases with exonuclease proofreading.

\section{Conclusion and perspectives}
\label{sec:Conclusion}

In this paper, the DNA multireplication process schematically representated in Fig.~\ref{fig1} was numerically simulated for five different DNA polymerases in order to test the validity of Chargaff's second parity rule.  Every replication proceeds by the template-directed elongation of the copy, binding one by one the monomeric units, i.e., the nucleotides forming base pairs with the template according to the biochemical kinetics of the DNA polymerase.  Many replications are successively performed to simulate the molecular evolution of the DNA sequence under the effect of the point-like mutations due to nucleotide substitutions by the polymerase according to its error probability~$\eta$.  After a number of replications of the order of the inverse $\eta^{-1}$ of the polymerase error probability, the fractions of mono- and oligonucleotides are observed to converge toward compliance with Chargaff's second parity rule, i.e., toward approximate equalities between the intrastrand fractions of complementary mono- and oligonucleotides.  These numerical observations are confirmed with the theory of the multireplication process of Fig.~\ref{fig1}.

For polymerase kinetics independent of the previously incorporated nucleotide, the copy forms a Bernoulli chain and the replication process is ruled by a four-by-four transition matrix, giving the evolution of the intrastrand nucleotide fractions during the process.  This transition matrix has a most remarkable structure:
\begin{itemize}
\item[(1)] its dominant part is a four-by-four anti-diagonal exchange matrix, expressing the fact that DNA replication would copy a nucleotide into the complementary nucleotide if there was no replication error;
\item[(2)] the rest of the transition matrix is of the order of the polymerase error probability $\eta$.
\end{itemize}
The multireplication process evolves by many iterations generated by matrix multiplication. Under this process, the intrastrand nucleotide fractions converge toward stationary values $(A,C,G,T)$ given by the eigenvector of the transition matrix that is associated with the leading eigenvalue equal to the unit value, $\lambda=1$.  These asymptotic intrastrand nucleotide fractions satisfy the equalities $A^{(0)}=T^{(0)}$ and $C^{(0)}=G^{(0)}$ at order $0$ of a series expansion in powers of the polymerase error probability $\eta$.  The equalities are no longer satisfied at higher orders $O(\eta^n)$: $A^{(n)}\ne T^{(n)}$ and $C^{(n)} \ne G^{(n)}$ for $n\ge 1$.  Therefore, theory demonstrates that the equalities between the asymptotic intrastrand fractions of complementary nucleotides are approximate with possible deviations of the order of the polymerase error probability, $\vert A-T\vert=O(\eta)$ and $\vert C-G\vert=O(\eta)$, which is in accordance with Chargaff's second parity rule.  The higher the fidelity $\eta^{-1}$ of the polymerase, the closer the equalities between the fractions of complementary nucleotides.  The fractions of oligonucleotides behave similarly.  The deviations with respect to the strict equalities $A^{(0)}=T^{(0)}$ and $C^{(0)}=G^{(0)}$ can also be explained in terms of differential equations ruling the evolution of intrastrand DNA composition every two replications over long enough time scales.

Furthermore, theory shows that the relaxation toward compliance with Chargaff's second parity rule is determined by the three eigenvalues beyond the leading one ($\lambda=1$), which have typically an absolute value smaller than the unit value by a quantity of the order of the error probability: $\vert\lambda\vert < 1$ and $1-\vert\lambda\vert =O(\eta)$.  Accordingly, the time scale of relaxation toward compliance (measured in number of replications) is of the order of the inverse of the polymerase error probability $\eta^{-1}$.  As a consequence, the higher the fidelity of the polymerase, the longer the relaxation toward compliance with Chargaff's second parity rule.  The relaxation time scale can be evaluated as the inverse error probability multiplied by the doubling time of organisms, which gives relatively short periods of time with respect to evolutionary time scales for unicellular organisms.

The results extend to more complicated polymerase kinetics having a dependence on the previously incorportated nucleotide.

In addition, the study also shows that the asymptotic nucleotide fractions depend on the kinetics of the polymerase and on the intracellular concentrations of the four nucleotides.  The latter are regulated by the nucleotide metabolism of the cell.  Interestingly, for typical polymerases and known physiological values of the nucleotide concentrations, the DNA nucleotide content is found to be more abundant in adenine and thymine than in cytosine and guanine, i.e., $A+T > C+G$, which corresponds to what is observed for the DNA of many species \cite{KRC68II,KRC68III,BHB02,OWS07,PNN14}.  

The mechanism, which is here theoretically studied, could be tested by automated experiments of multiple DNA replications successively performed by some DNA polymerase starting from some initial DNA sequence.

We emphasize that the model, which is used as the vehicle of this study, is far from realistic.  In general, replication is carried out by several types of polymerases and the intracellular nucleotide concentrations may vary during the different phases of cell division. Accordingly, the expectation is that the DNA nucleotide fractions should result from an average over the different steps of replication.  Nevertheless, all the polymerases replicate every DNA strand into a mostly complementary strand, so that the main features of replication are essentially captured by the model of Fig.~\ref{fig1}.  In this regard, the replication process based on the biochemical kinetics of polymerases provides a robust mechanism underlying the approximate intrastrand symmetry expressed by Chargaff's second parity rule.  The polymerase kinetics generate point-like mutations caused by replication errors, which drives the molecular evolution of DNA sequences toward compliance with Chargaff's second parity rule.  There is an evolutionary advantage for the genome to remain close to compliance.  Otherwise, the genome would drift away from its selected composition because of mutational pressure.

Finally, we remark that the robustness of the mechanism based on the kinetics of DNA polymerases suggests that the strong deviations with respect to Chargaff's second parity rule, which are observed in the genome of mammalian mitochondria and in some single-stranded DNA viruses \cite{MB06,NA06}, might be caused by the interference of DNA replication with other processes such as transcription \cite{YRCYBJH06} in combination with the smallness of these genomes.  Future studies should address this open issue.

\begin{acknowledgments}
The author thanks Thomas Gilbert and Jean Lobry for fruitful discussions.
This research is supported by the Universit\'e libre de Bruxelles (ULB).
\end{acknowledgments}

\appendix

\section{Mathematical formulation of replication kinetics}
\label{app:simul-B}

\subsection{Kinetic equations}

The replication process~(\ref{reaction}) is ruled by the following kinetic equations \cite{G16a},
\bea
\frac{d}{dt}\, P_t\left(m_1 \cdots m_l \qquad\quad\ \atop n_1\, \cdots \, n_l \, n_{l+1}\cdots\right) &=& 
W_{+m_l,l} \, P_t\left(m_1 \cdots m_{l-1} \qquad\ \ \atop n_1\, \cdots \, n_{l-1} \, n_l\cdots\right) \nonumber\\
&+&\sum_{m_{l+1}} W_{-m_{l+1}, l+1}
\, P_t\left(m_1 \cdots m_l m_{l+1} \qquad\ \ \atop n_1\, \cdots \, n_l \, n_{l+1}\, n_{l+2}\cdots\right)
\nonumber\\
&-& \left( W_{-m_l,l}
+\sum_{m_{l+1}} W_{+m_{l+1},l+1}\right)  
P_t\left(m_1 \cdots m_l \qquad\quad \ \atop n_1\, \cdots \, n_l \, n_{l+1}\cdots\right) ,
\label{kin_eq}
\eea
for the probabilities of all the copy sequences $m_1$, $m_1m_2$, $\dots$, $m_1 \cdots m_l $, $\dots$, which may be generated along the template sequence $n_1\, \cdots \, n_l \, n_{l+1}\cdots$.  In these kinetic equations, $W_{\pm m_l,l}$ denote, respectively, the attachment and detachment rates of the nucleotide $m_l$ at the location $l$ of the template.  The polymerases are enzymes, which basically obey Michaelis-Menten kinetics \cite{MM13,S81,L82}.  The mathematical expressions of the rates are known whether the kinetics of the polymerase depend or not on the previously incorporated nucleotide \cite{G16a,G16PRL,G16PTRSA,G17PRE}.

\subsection{Kinetics independent of previously incorporated nucleotide}

For kinetics independent of the previously incorporated nucleotide (as considered in Sections~\ref{sec:simul-B} and~\ref{sec:math-theory}), the attachment and detachment rates can be expressed in terms of the constants $k^{\rm p}_{\pm m_{l}\atop \ \, n_{l}}$ for the polymerization and depolymerization of $m_{l}$ after its pairing with $n_{l}$ and the dissociation constant $K_{m_{l}\atop n_{l}}$ of the pair $m_{l}\!:\!n_{l}$.  Moreover, these rates depend on the concentrations $[m_l{\rm PP}]$ of the deoxyribonucleoside triphosphates $m_l{\rm PP}$ and on the concentration~$[{\rm PP}]$ of pyrophosphate~${\rm PP}$. The concentration unit is the {\it mole per liter}  (M).  With these notations, the attachment rate of $m_{l}$ is given by
\be
W_{+m_l,l} = W_{+m_{l}\atop \ \, n_{l}} =  
\frac{k^{\rm p}_{+m_{l}\atop \ \, n_{l}}[m_{l}{\rm PP}]}
{K_{m_{l}\atop n_{l}}\, Q_{\atop n_{l}}}
\label{W+B}
\ee
and the detachment rate of $m_l$ by
\be
W_{-m_l,l} = W_{\quad\, -m_l\atop n_{l+1}\, n_l} = 
\frac{k^{\rm p}_{-m_l\atop \ \, n_l}[{\rm PP}]}{Q_{\atop n_{l+1}}} \, ,
\label{W-B}
\ee
where
\be
Q_{\atop n_{l}}\equiv 1 + \sum_{m_{l}} \frac{[m_{l}{\rm PP}]}{K_{m_{l}\atop n_{l}}}
\label{denom}
\ee
is the Michaelis-Menten denominator \cite{G16a}.  We note that the denominator in the depolymerization rate~(\ref{W-B}) is shifted by one unit with respect to the denominator in the polymerization rate~(\ref{W+B}) because the rapid intermediate step of the Michaelis-Menten kinetics is the formation of the pair $m_l\!:\!n_l$ for the forward reaction, but that of the pair $m_{l+1}\!:\!n_{l+1}$ for the backward reaction \cite{G16a}.

Experimental data being rare for the depolymerization rate constants, it is assumed for simplicity that they are proportional to the corresponding polymerization rate constant according to
\be
k^{\rm p}_{-m_{l}\atop \ \, n_{l}}= \frac{1}{K_{\rm P}} \, k^{\rm p}_{+m_{l}\atop \ \, n_{l}} \, ,
\label{K_P-dfn}
\ee
where the constant associated with pyrophosphorolysis is here supposed to take the value $K_{\rm P}=200$~mM \cite{G16a}.  The physiological concentration of pyrophosphate is typically given by $[{\rm PP}]=10^{-4}$~M \cite{H01}.

\subsection{Kinetics with a dependence on previously incorporated nucleotide}
\label{app:A3}

For such kinetics, the constants also depend on the penultimate pair $m_{l-1}\!:\!n_{l-1}$ \cite{G16PRL,G16PTRSA,G17JSM,G17PRE}.  Accordingly, the attachment and detachment rates are given by similar expressions as Eqs.~(\ref{W+B})-(\ref{denom}), but with the constants $k^{\rm p}_{\pm m_{l}\atop \ \, n_{l}}$ and $K_{m_{l}\atop n_{l}}$ replaced with $k^{\rm p}_{{\pm m_{l}\atop \ \, n_{l}}\!\big\vert\!{m_{l-1}\atop n_{l-1}}}$ and $K_{{m_{l}\atop \ \, n_{l}}\!\big\vert\!{m_{l-1}\atop n_{l-1}}}$, respectively.

\section{Differential equations for continuous-time evolution}
\label{app:ODEs}

Every two replications, the evolution of nucleotide fractions according to Eq.~(\ref{next-p_r}) is smooth enough to be well approximated by a continuous-time evolution process ruled by linear differential equations obtained from
\be
{\bf p}(t+2\tau)-{\bf p}(t) = 2\tau \frac{d{\bf p}(t)}{dt} + O(\tau^2) = \left(\boldsymbol{\mathsf P}^2 - \boldsymbol{\mathsf 1}\right)\cdot{\bf p}(t) \, ,
\label{differential}
\ee
where $\tau$ is the time step due to one replication.  Since the $4\times 4$ transition matrix~(\ref{matrix-P}) has the form $\boldsymbol{\mathsf P}=\boldsymbol{\mathsf C}+\boldsymbol{\mathsf P}^{(1)}$, where $\boldsymbol{\mathsf C}$ is the antidiagonal exchange matrix~(\ref{P0=C}), we find that
\be
\boldsymbol{\mathsf P}^2 = \boldsymbol{\mathsf 1} + \boldsymbol{\mathsf C}\cdot\boldsymbol{\mathsf P}^{(1)}+\boldsymbol{\mathsf P}^{(1)}\cdot\boldsymbol{\mathsf C} + \left( \boldsymbol{\mathsf P}^{(1)} \right)^2 \, ,
\ee
as a result of the property $\boldsymbol{\mathsf C}^2=\boldsymbol{\mathsf 1}$.  Because the square of the transition matrix $\boldsymbol{\mathsf P}^2$ has eigenvalues $\lambda^2$ that are equal or very close to the unit value [i.e., because $\rho=(-\ln\vert\lambda\vert)^{-1}=O(\eta^{-1}) \gg 1$], the term $O(\tau^2)$ is negligible in Eq.~(\ref{differential}).  For this reason, over long time scales $t\gg\tau$, the nucleotide fractions obey the following set of differential equations,
\be
\label{dpdt=Mp}
\frac{d{\bf p}}{dt} = \boldsymbol{\mathsf M}\cdot{\bf p} \, ,
\qquad\mbox{where}\qquad
\boldsymbol{\mathsf M}=\boldsymbol{\mathsf M}^{(1)}+\boldsymbol{\mathsf M}^{(2)}
\ee
with
\be
\boldsymbol{\mathsf M}^{(1)} \equiv \frac{1}{2\tau} \left(\boldsymbol{\mathsf C}\cdot\boldsymbol{\mathsf P}^{(1)}+\boldsymbol{\mathsf P}^{(1)}\cdot\boldsymbol{\mathsf C}\right) 
\label{M1}
\ee
and
\be
\boldsymbol{\mathsf M}^{(2)} \equiv \frac{1}{2\tau} \left( \boldsymbol{\mathsf P}^{(1)} \right)^2 \, . 
\label{M2}
\ee
Since $\boldsymbol{\mathsf P}^{(1)}=O(\eta)$, we have that $\boldsymbol{\mathsf M}^{(1)}=O(\eta)$ and $\boldsymbol{\mathsf M}^{(2)}=O(\eta^2)$.

As a direct consequence of the special structure of the matrix $\boldsymbol{\mathsf C}$, the leading matrix~(\ref{M1}) is given by
\be
\boldsymbol{\mathsf M}^{(1)} = 
\left[
\begin{array}{cccc}
-a-c-e & d & b & a \\
e & -b-d-f & f & c \\
c & f & -b-d-f & e \\
a & b & d & -a-c-e
\end{array}
\right] ,
\label{M1=4x4}
\ee
which only depends on the following six coefficients,
\bea
&& a \equiv \frac{1}{2\tau} \left( P_{{\rm A}\atop{\rm A}} + P_{{\rm T}\atop{\rm T}} \right) , \qquad
b \equiv \frac{1}{2\tau} \left( P_{{\rm A}\atop{\rm C}} + P_{{\rm T}\atop{\rm G}} \right) , \qquad
c \equiv \frac{1}{2\tau} \left( P_{{\rm C}\atop{\rm A}} + P_{{\rm G}\atop{\rm T}} \right) , \nonumber\\
&& d \equiv \frac{1}{2\tau} \left( P_{{\rm A}\atop{\rm G}} + P_{{\rm T}\atop{\rm C}} \right) , \qquad
e \equiv \frac{1}{2\tau} \left( P_{{\rm G}\atop{\rm A}} + P_{{\rm C}\atop{\rm T}} \right) , \qquad
f \equiv \frac{1}{2\tau} \left( P_{{\rm C}\atop{\rm C}} + P_{{\rm G}\atop{\rm G}} \right) ,
\label{6-coeff}
\eea
although the transition matrix~(\ref{matrix-P}) has twelve independent coefficients.
The matrix~(\ref{M1=4x4}) has precisely the same form as in the six-parameter model based on the no-strand-bias conditions~\cite{S95,L95,LL99}. This matrix has the following symmetry,
\be
\boldsymbol{\mathsf C}^{-1}\cdot\boldsymbol{\mathsf M}^{(1)} \cdot\boldsymbol{\mathsf C} = \boldsymbol{\mathsf M}^{(1)} \, ,
\ee
because of $\boldsymbol{\mathsf C}^2=\boldsymbol{\mathsf 1}$ and $\boldsymbol{\mathsf C}^{-1}=\boldsymbol{\mathsf C}$.

However, there exist deviations caused by the matrix~(\ref{M2}), which breaks the symmetry
\be
\boldsymbol{\mathsf C}^{-1}\cdot\boldsymbol{\mathsf M}^{(2)} \cdot\boldsymbol{\mathsf C} \ne \boldsymbol{\mathsf M}^{(2)}
\ee
at next order in the polymerase error probability.  The smallness of $\boldsymbol{\mathsf M}^{(2)}$ explains that this second-order contribution may go unnoticed.

The solutions of the set of differential equations~(\ref{dpdt=Mp}) converge toward the stationary solution given by the eigenvector~$\bf v$ satisfying Eq.~(\ref{eq-v}), which implies the stationarity condition $\boldsymbol{\mathsf M}\cdot{\bf v}=0$.  This stationary solution can be expanded according to Eq.~(\ref{v-series}).  The leading term of this expansion is given by Eq.~(\ref{v(0)}) with
\be
A^{(0)} = T^{(0)} = \frac{b+d}{2(b+c+d+e)} 
\qquad\mbox{and}\qquad
C^{(0)} = G^{(0)} = \frac{c+e}{2(b+c+d+e)} \, ,
\ee
which agrees with the formulas obtained in Refs.~\cite{L95,LL99} and which leads to the formulas~(\ref{A0=T0})-(\ref{C0=G0}) because of Eq.~(\ref{6-coeff}).  Therefore, strict equalities are given by the stationarity equation $\boldsymbol{\mathsf M}^{(1)}\cdot{\bf v}^{(0)}=0$ at leading order, although this is not the case for the solution $\bf v$ of the full equation $\boldsymbol{\mathsf M}\cdot{\bf v}=0$, for which $A\ne T$ and $C\ne G$.

These results show that there are significant similarities and differences between the mechanism described in Section~\ref{sec:math-theory} and the model based on the no-strand-bias conditions \cite{S95,L95,LL99}.



\end{document}